# Response of the lattice across the filling-controlled Mott metal-insulator transition of a rare earth titanate

Honggyu Kim [a)], Patrick B. Marshall, Kaveh Ahadi, Thomas E. Mates, Evgeny Mikheev [b)], and Susanne Stemmer [c)]

*Materials Department, University of California, Santa Barbara, CA 93106-5050, USA*

[a)] Electronic mail: kyuya380@gmail.com

[b)] Present address: SLAC National Accelerator Laboratory, MC 4045, Stanford University, Stanford, California 94305, USA.

[c)] Electronic mail: stemmer@mrl.ucsb.edu

**Abstract**

The lattice response of a prototype Mott insulator, SmTiO$_3$, to hole doping is investigated with atomic-scale spatial resolution. SmTiO$_3$ films are doped with Sr on the Sm site with concentrations that span the insulating and metallic sides of the filling-controlled Mott metal-insulator transition (MIT). The GdFeO$_3$-type distortions are investigated using an atomic resolution scanning transmission electron microscopy technique that can resolve small lattice distortions with picometer precision. We show that these distortions are gradually and uniformly reduced as the Sr concentration is increased without any phase separation. Significant distortions persist into the metallic state. The results present a new picture of the physics of this prototype filling-controlled MIT, which is discussed.

The Mott metal-insulator transition (MIT) remains one of the most complex problems in condensed matter physics [1]. Ideally, the Mott insulating state is due to on-site electron-electron repulsions that localize the conduction electrons of a narrow, half-filled band and open up an energy gap that is determined by the Coulomb repulsion energy, $U$ [2, 3]. In this picture, the Mott insulator exists only at half-filling. While many important Mott insulators, such as the cuprate superconductors [4], exhibit "filling-controlled" transitions to a metal, they often require substantial amounts of doping (a few % or more), which is inconsistent with the ideal picture. A related question concerns the pathways between the metal and the insulator: how do mobile electrons become localized spins [5]? Changes in the crystal symmetry often accompany the MIT in real materials. This makes it difficult to distinguish the respective roles of very different types of interactions, such as electron-electron and electron-lattice [6]. Many symmetry-breaking MITs are first order and result in metal/insulator phase coexistence near the transition [7-10]. In contrast, continuous, second order transitions can be quantum critical and, as a result, are not strictly tied to a specific materials symmetry [11-13]. They give rise to truly novel (non-Fermi liquid) quasi-particle behavior [14, 15].

Rare-earth titanates ($R$TiO$_3$, where $R$ is a rare earth ion, but not Eu) are prototype Mott insulators with a $d^1$ electron configuration. Coulomb repulsion splits a half-filled 3$d$ conduction band into upper and lower Hubbard bands, respectively. The Mott-Hubbard gap collapses upon hole doping [16-18]. Their relative structural and chemical simplicity makes the $R$TiO$_3$s nearly ideal systems for understanding the microscopic mechanisms of a filling-controlled Mott MIT in real materials.

It is well established that the orthorhombic GdFeO$_3$-type distortions (*Pbnm* space group) [19], which vary systematically with the size of the $R$-ion, determine the electronic structure of the $R$TiO$_3$s [20-22]. The required doping concentration for this filling-controlled transition scales with the magnitude of the GdFeO$_3$-type distortions [23]. Some studies suggest evidence for phase coexistence near this filling-controlled transition [17], which may imply an underlying, first-order, *bandwidth-controlled* MIT [24, 25]. Other studies, however, suggest a quantum critical point [26], which implies a continuous, second order transition.

Here, we show that by correlating the atomic scale structure with the electrical properties across the filling-controlled MIT of SmTiO$_3$ new insights into this prototype Mott MIT can be obtained. We characterize the local lattice response, using atomic resolution scanning transmission electron microscopy (STEM), which can resolve very small distortions, including those around individual point defects [27-29]. This allows us to study the response of the lattice to hole (Sr) doping on length scales that range from extremely local (i.e., the distortion of atomic columns surrounding the dopant) to long-range (by analyzing images from different regions). We find that this MIT is not strongly coupled to a specific lattice symmetry. The results also provide evidence for a continuous phase transition with no phase separation. They suggest an electron correlation-driven transition, to which the lattice responds. This is further corroborated by a surprisingly non-local response of the lattice to the presence of the Sr dopant atoms.

Sm$_{1-x}$Sr$_x$TiO$_3$ thin films were grown by molecular beam epitaxy (MBE) on (001) (La$_{0.3}$Sr$_{0.7}$)(Al$_{0.65}$Ta$_{0.35}$)O$_3$ (LSAT). Details of the film growth can be found elsewhere [30]. The Sr (hole) doping concentrations ($x$) ranged from 0 to 0.2, spanning the MIT, which occurs near $x \sim 0.05$. Sr concentrations were estimated from the beam flux ratios during growth, which provide a reasonable estimate because of the high sticking coefficients of Sr and Sm, and verified using ex-situ x-ray photoelectron spectroscopy (see Supplementary Information [31]). Both measurements were found to be in good agreement. Electrical measurements were carried out using a Physical Property Measurement System (Quantum Design PPMS) using four-point probe, Van-der Pauw geometry. Figure 1(a) shows the sheet resistances ($R_s$) as a function of temperature for films with different amounts of Sr ($x$), which change systematically with $x$. SmTiO$_3$ films were insulating with resistances exceeding the measurement limit and are therefore not shown in Fig. 1. We investigated films with different thicknesses (30 – 50 nm) and found no significant differences in the transport behavior. The Sm$_{0.95}$Sr$_{0.05}$TiO$_3$ film is at the MIT boundary, showing only a small variation of the resistance with temperature, whereas films with $x \geq 0.1$ are metallic (d$R_s$/d$T$ > 0). The temperature behavior and values of $R_s$ as a function of Sr doping are similar to those reported in the literature for the other $R$TiO$_3$s [1].

For STEM, cross-section TEM samples of films with $x$ = 0, 0.05, and 0.10 were prepared using a FEI Helios Dualbean Nanolab 650 focused ion beam system with 2 keV Ga ions. High-angle annular dark-field (HAADF) imaging in STEM was carried out at room temperature using a 300 kV FEI Titan S/TEM ($C_s$ = 1.2 mm) with a convergence semi-angle of 9 mrad and a HAADF detector range and 60 mrad - 390 mrad. To enhance the signal-to-noise ratio and reduce the effect of scan distortions, 20 images (1024×1024 pixels, 1 s frame time) were sequentially recorded, rigidly aligned, and then averaged using a cross-correlation method, from which the position of atomic columns are obtained by a two-dimensional (2D) Gaussian peak fitting. This approach allows for measurements of displacements with picometer precision, as described in detail elsewhere [28].

SmTiO$_3$ films grow on LSAT with an epitaxial orientation relationship described by (110)$_o$||(001)$_c$ (the subscripts refer to the orthorhombic-like unit cell for the film and the cubic substrate, respectively), which minimizes the lattice mismatch. The two possible orientation domains, (110)$_o$||(010)$_c$ and (001)$_o$||(010)$_c$, are shown in Figs. 2(a) and 2(b), respectively. Figures

2(c) and 2(d) show HAADF images of SmTiO$_3$ imaged along [110]$_o$ and [001]$_o$, respectively. Because of the atomic number (Z) sensitivity of HAADF, Sm columns appear brightest. In the GdFeO$_3$ structure, the A-site cations displace proportionally to the degree of TiO$_6$ octahedra rotations and thus provide a measure of the orthorhombic distortions [34, 35]. Sm displacements can be discerned along [110]$_o$, but not along [001]$_o$. Using the HAADF images recorded along [110]$_o$, we quantify the orthorhombic-like distortions using two "deviation angles", (180° - $\theta_{xx}$) and (90° - $\theta_{xy}$), defined in Fig. 2(c), which are determined by three successive Sm atomic columns. The angle (180° - $\theta_{xx}$) of the SmTiO$_3$ film is 14.40 ± 1.72°, which agrees well with the expected value (14.7°) for a strained film on LSAT, as calculated from the SmTiO$_3$ crystal structure [36]. This shows that the structure is bulk-like, despite the thin TEM sample.

Figures 3(a) and 3(b) show HAADF images of Sm$_{1-x}$Sr$_x$TiO$_3$ films with $x$ = 0.05 and 0.10, respectively. The average deviation angles (180° - $\theta_{xx}$), obtained by from different regions with a total sample area of ~ 320 nm$^2$, are 11.90° ± 1.67° for $x$ = 0.05 and 9.98° ± 1.76° for $x$ = 0.1. Thus, the orthorhombic-like distortions decrease with increasing amount of Sr. Even the metallic film, however, retains a sizeable orthorhombic-like distortion; it is not cubic (for cubic, 180°- $\theta_{xx}$ = 0°, which is very different from 180°- $\theta_{xx}$ = 9.98° in the metallic film). For cubic films, STEM measures 180°- $\theta_{xx}$ < 1.5°, see ref. [37]).

Figure 3(c) shows the statistical distributions of $\theta_{xx}$ for the three films, which determines the errors given above. The average unit cell distortions gradually decrease with Sr content and the standard deviations are similar to that of the undoped SmTiO$_3$ film. The standard deviation of the SmTiO$_3$ film reflects the measurement error. *The constant standard deviation for all three films shows that the decrease in unit cell distortions with Sr doping proceeds in a spatially uniform manner.*

To clarify this statement, which has significant impact on the nature of the MIT, we note that two types of structural inhomogeneities may have occurred with Sr doping: (i) Phase separation into cubic (metallic) and orthorhombic-like (insulating) regions, if the MIT is first order. (ii) Local displacements around the Sr dopants. Such local atom relaxations typically occur around point defects in perovskite oxides, due to a variety of factors, such as size and charge mismatch and a propensity for small polaron formation. The method used in the present study is capable of observing them, see refs. [27, 28]. Neither type of structural non-uniformity are observed here, as will be discussed next.

Starting with (i), the constant standard deviations, which were obtained across images recorded from several different regions of each sample, show that upon approaching the MIT, no phase separation into cubic (or less distorted) and orthorhombic regions occurs. We emphasize that this result is not a projection issue through the finite thickness of the TEM sample (~10 nm – 20 nm, see Supplementary Information). First, we emphasize that the method employed here *can* detect atomic-scale variations in local displacements, when present, even in cases where they affect *only* a few atoms, with picometer precision, despite the fact that along the projection direction through a finite TEM sample thickness not all the atoms in a column may be displaced, see refs. [27, 28]. Therefore, even very small regions (few nm) that have different distortions should be discernable. The experiments are clearly sufficiently sensitive to detect the differences in distortions of the metallic film and the insulating film, respectively. In case of phase separation, the differences would be even larger ($\Delta\theta_{xx} > 10°$) and thus should be reflected in the standard deviation of $\theta_{xx}$. Further confirmation comes from the fact that the standard deviation was found to be independent of the TEM sample thickness (see Supplementary Information [31]). Therefore, no structural phase separation occurs in these samples.

To further examine (ii), we measured atomic column intensities ($I_{Sm,Sr}$) and the deviation angles (90° - $\theta_{xy}$), which reflect the chemical content of the atomic columns. Figure 4 shows HAADF images (left), $I_{Sm,Sr}$ intensity maps (middle) and (90° - $\theta_{xy}$) maps (right) for films with $x$ = 0.05 and 0.10, respectively. The variations in $I_{Sm,Sr}$ across the image are due to thickness gradients of the TEM sample, i.e., the thickness decreases from the bottom towards top region in both images. Another source of $I_{Sm,Sr}$ variations is the Sr doping. The Sr dopants have a lower $Z$ than the host (Sm); thus, Sr-containing columns show lower image intensity, with the column intensity depending on the number of Sr atoms, their depth locations, the total number of atoms in the column, and the STEM parameters [28, 38-40]. The white arrows in the $I_{Sm,Sr}$ maps indicate selected columns that show significantly reduced intensities relative to those of nearest-neighbor Ti-O columns, which have similar thicknesses. We note that a significant fraction of columns in each image do not contain Sr dopants (see Supplementary Information [31]). The corresponding deviation angle (90° - $\theta_{xy}$) maps exhibit the modulation of positive (red) and negative (blue) values of the structure. The different degrees of distortions of the two films are visible (note the differences in color ranges). Interestingly, unlike the intensity maps, the deviation angles remain uniform, even for those columns that contained Sr (white arrows). This result shows that, surprisingly, despite the size and nominal ionic charge mismatch, the Sr dopants do not locally distort the lattice, within the picometer sensitivity of the method.

To summarize, the results show that Sr doping gradually and uniformly changes the degree of octahedral distortions without any structural phase separation. The Sr atoms do not locally affect the neighboring atoms but instead produce a *global structure change*. Furthermore, there is no abrupt symmetry change as the MIT boundary is traversed with doping.

The findings impact the understanding of this prototype filling-controlled Mott transition. First, the transition is not first order, because we do not observe phase separation into regions with distinguishable symmetries. At present, we cannot exclude a purely electronic phase separation, although there would be a large electrostatic penalty for an electronic phase separation that is decoupled from structural nonuniformity. Secondly, considerable GdFeO$_3$-like lattice distortions persist into the metallic phase. The structure changes gradually across the transition while the electrical properties change by a large amount. This implies that this filling-controlled MIT is *not* coupled to a particular symmetry of lattice. Interestingly, our prior studies have shown that the magnetic properties are also relatively independent of the octahedral distortions [35]. The behavior can be contrasted with materials such as the rare earth nickelates, which show a temperature-driven, first order transition that is tightly coupled to a specific symmetry of the insulator that allows for charge order [41, 42]. The results support a picture of the MIT in the $R$TiO$_3$s that is driven by electron-electron correlations that remain important into the metallic phase.

The question remains as to why large hole doping concentrations (~5%) are needed to induce the transition. We can rule out a percolative, first-order transition as the explanation [24, 25]. A possible explanation involves disorder introduced by the Sr dopants and which causes the carriers to localize, i.e. the MIT is of Mott-Anderson type [43]. The results also show, however, that the Sr dopants have a global, long-range effect on the structure and that the lattice responds even on the metallic side of the transition. The behavior of the Hall coefficient ($R_H$) in the metallic phase appears to corroborate the idea that a simple doping picture does not apply in this system. In such a picture, $(eR_H)^{-1}$, where $e$ is the elementary charge, should correspond to the dopant concentration. As can be seen from Fig. 1(b), this is true for SrTiO$_3$-rich compositions, but $(eR_H)^{-1}$ strongly deviates from the expected value [line in Fig. 1(b)] in metallic films with $x <$ 0.7, suggesting the onset of significant changes in the electronic structure, along with the global response of the lattice, and a collective electronic state already far from the MIT. The deviation increases as the MIT is approached ($x \sim 0.05$). We note that in Sr-doped LaTiO$_3$, the deviation of $(eR_H)^{-1}$ from nominal doping is also observed, but occurs closer to the MIT boundary [16]. Other doped Mott insulators, such as the cuprate superconductors, exhibit complex behavior of $(eR_H)^{-1}$ as a function of doping (see e.g. ref. [44]).

To conclude, the results show that this prototype filling-controlled Mott MIT is of second order and not strongly coupled to a specific lattice symmetry. Furthermore, we have shown that this doped Mott insulator presents a radically different picture expected from a conventional dopant, both electrically and in terms of the lattice response. Future experimental studies of the evolution of the electronic structure, for example by scanning tunneling microscopy or angle resolved photoemission, may shed light on how the collective electronic response triggers the global, spatially uniform response of the lattice that persists into the metallic state.

**Acknowledgements**

The authors thank Jim Allen and Leon Balents for helpful discussions. The work was supported by the U.S. Department of Energy (Grant No. DEFG02-02ER45994) and by a MURI funded by the U.S. Army Research Office (Grant No. W911NF-16-1-0361).

**Figure Captions**

**Figure 1:** (a) Temperature dependence of the sheet resistance for 50-nm-thick $Sm_{1-x}Sr_xTiO_3$ films with $x$ = 0.05, 0.10, 0.15, and 0.20. (b) Apparent carrier density, $(eR_H)^{-1}$, measured from the Hall effect for $Sm_{1-x}Sr_xTiO_3$ films with different compositions, with the right most data points from the series that as investigated in the present study. The solid line indicates the expected values for the composition for a simple metal. The film thicknesses were 30 nm and 50 nm, respectively.

**Figure 2:** (a-b) Schematics showing two possible orientation variants of $SmTiO_3$ films on LSAT (green color): (a) $(110)_O \| (010)_C$ and (b) $(001)_O \| (001)_C$, respectively. (c-d) HAADF images of $SmTiO_3$ films grown on LAST recorded along (c) $[110]_O$ and (d) $[001]_O$, respectively. Two angles, $\theta_{xx}$ and $\theta_{xy}$, which are used to quantify the orthorhombic-like unit cell distortions, are shown in (c).

**Figure 3:** HAADF images of $Sm_{1-x}Sr_xTiO_3$ films with Sr concentrations of (a) 5 and (b) 10 %, respectively. (c) Distributions of measured deviation angles, $(180° - \theta_{xx})$, of $SmTiO_3$ and $Sm_{1-x}Sr_xTiO_3$ films show the reduction in of orthorhombic-like unit cell distortions with increasing $x$. The vertical axis (counts) are the number of measured deviation angles in all analyzed images.

**Figure 4:** HAADF images (left), $I_{Sm,Sr}$ maps (middle) and deviation angle $(90° - \theta_{xy})$ maps (right) of $Sm_{1-x}Sr_xTiO_3$ films with x = 0.05 (top row) and 0.1 (bottom row). The contrast of intensity maps is adjusted to maximize the intensity distribution. Low atomic column intensities marked by the white arrows in the $I_{Sm,Sr}$ maps and are associated with the atomic columns

containing Sr. Note that the pixel sizes of the intensity map and the deviation angle maps are different.

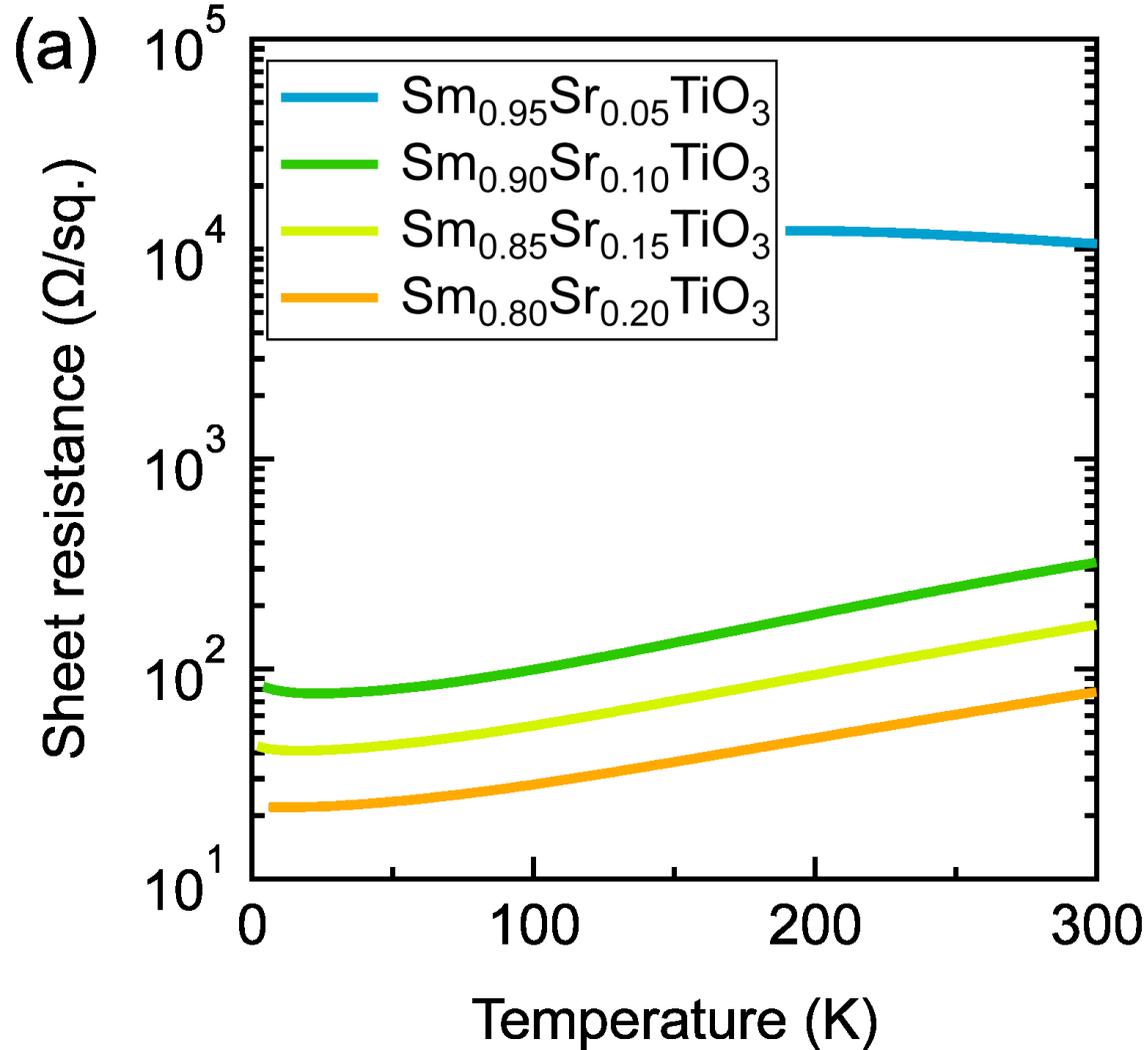

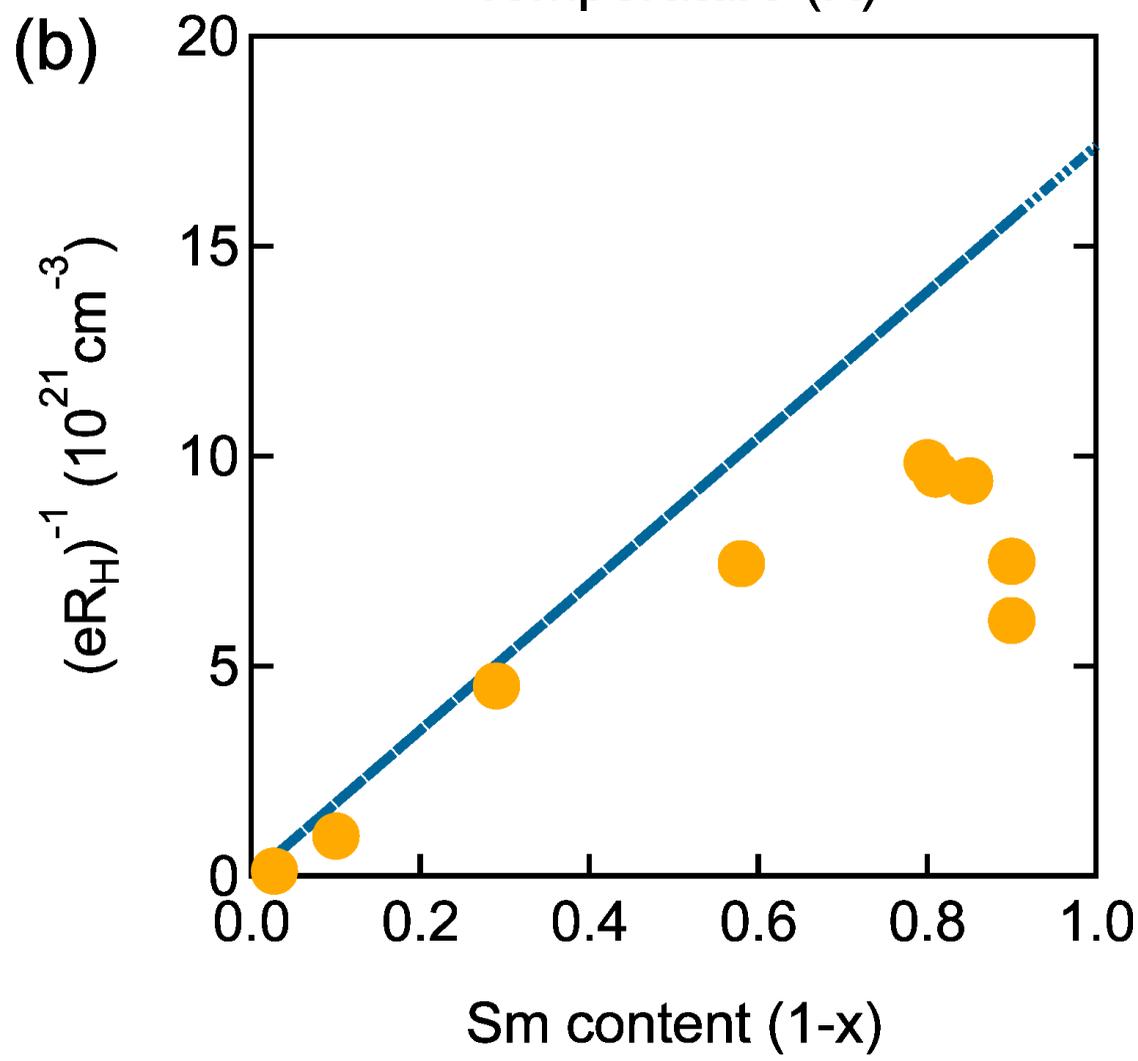

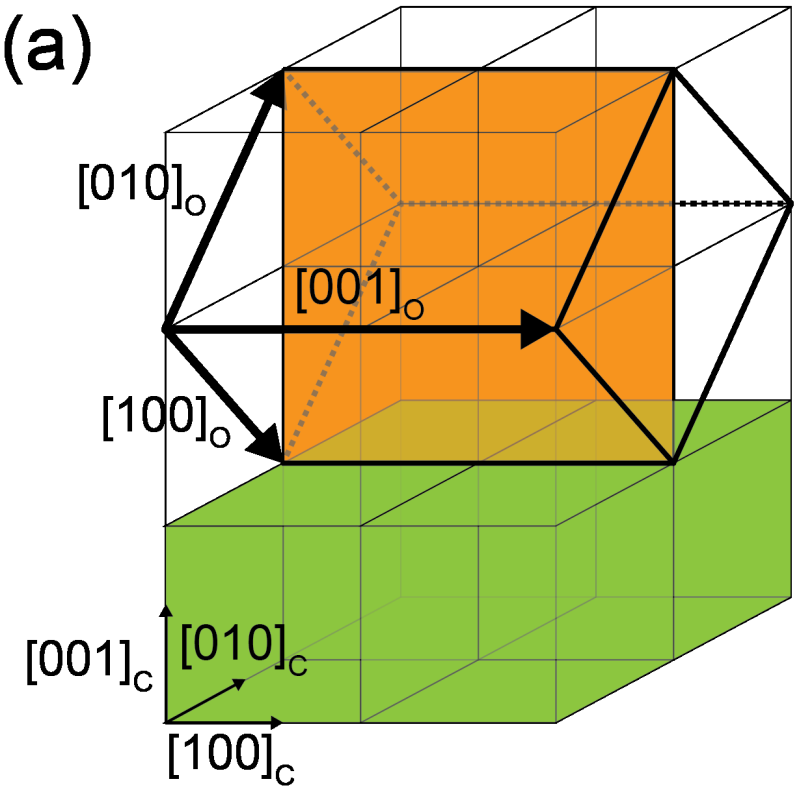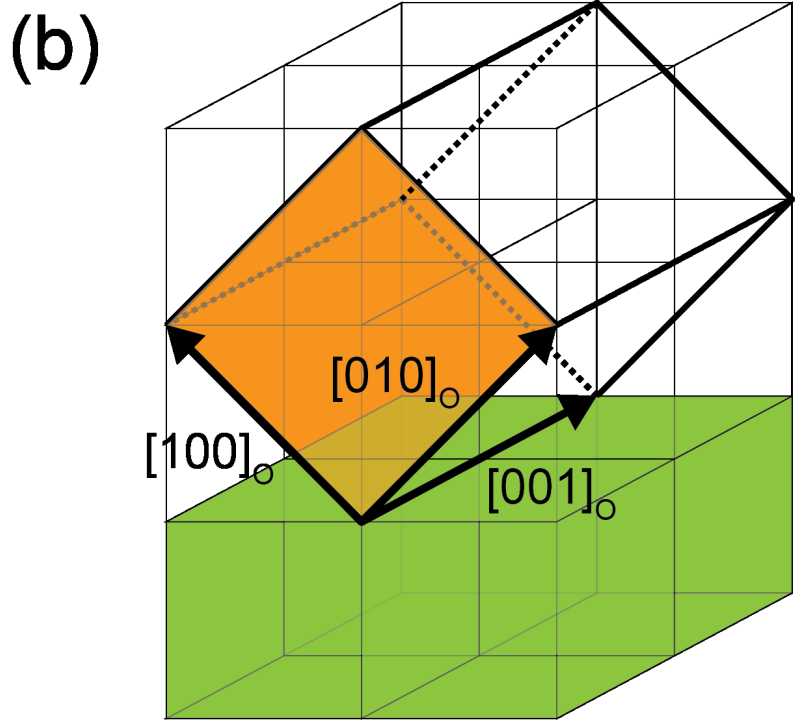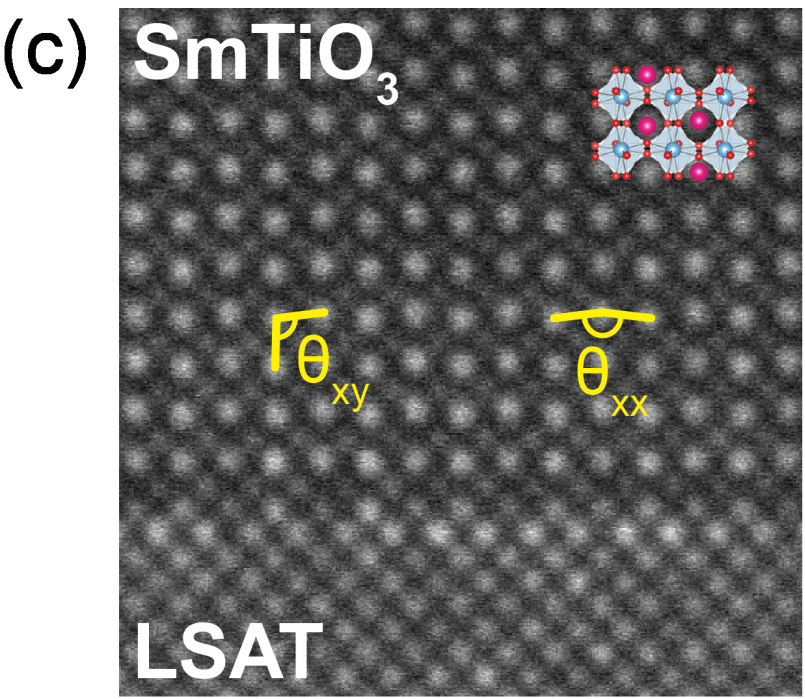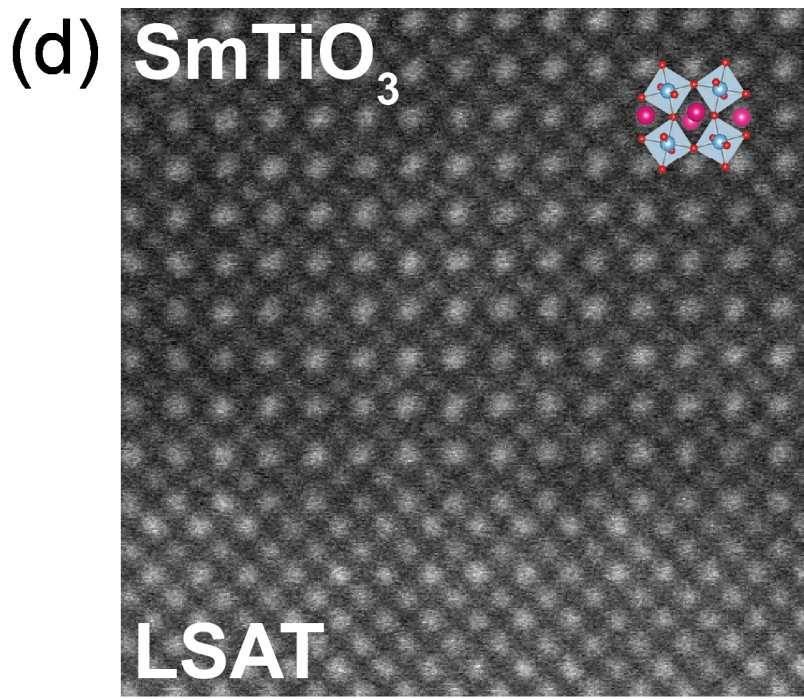

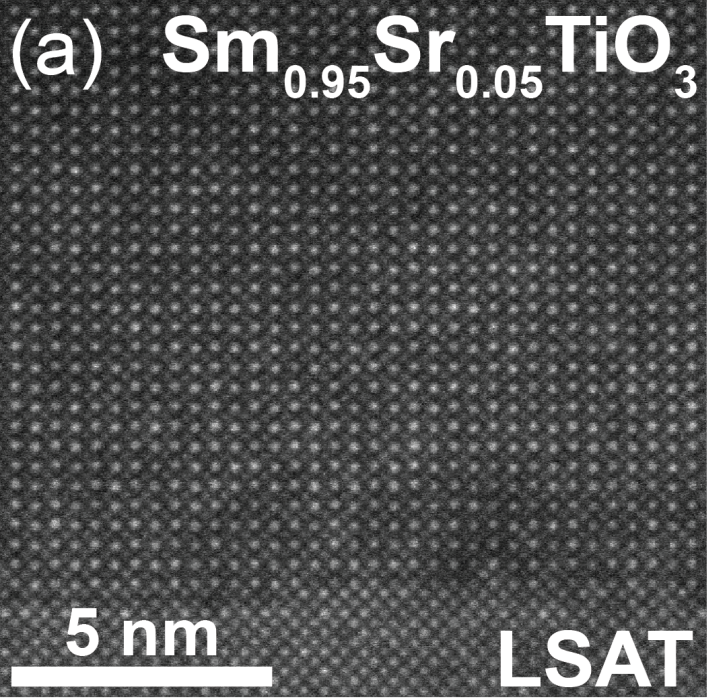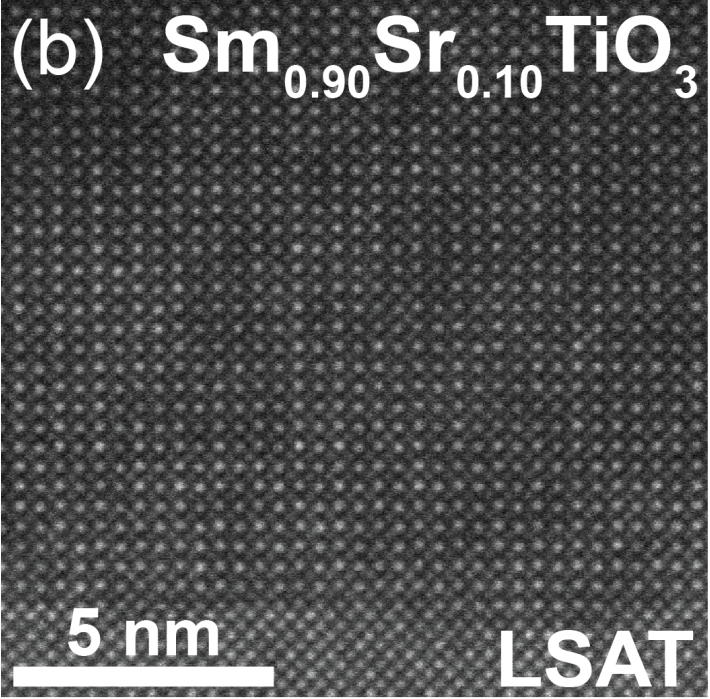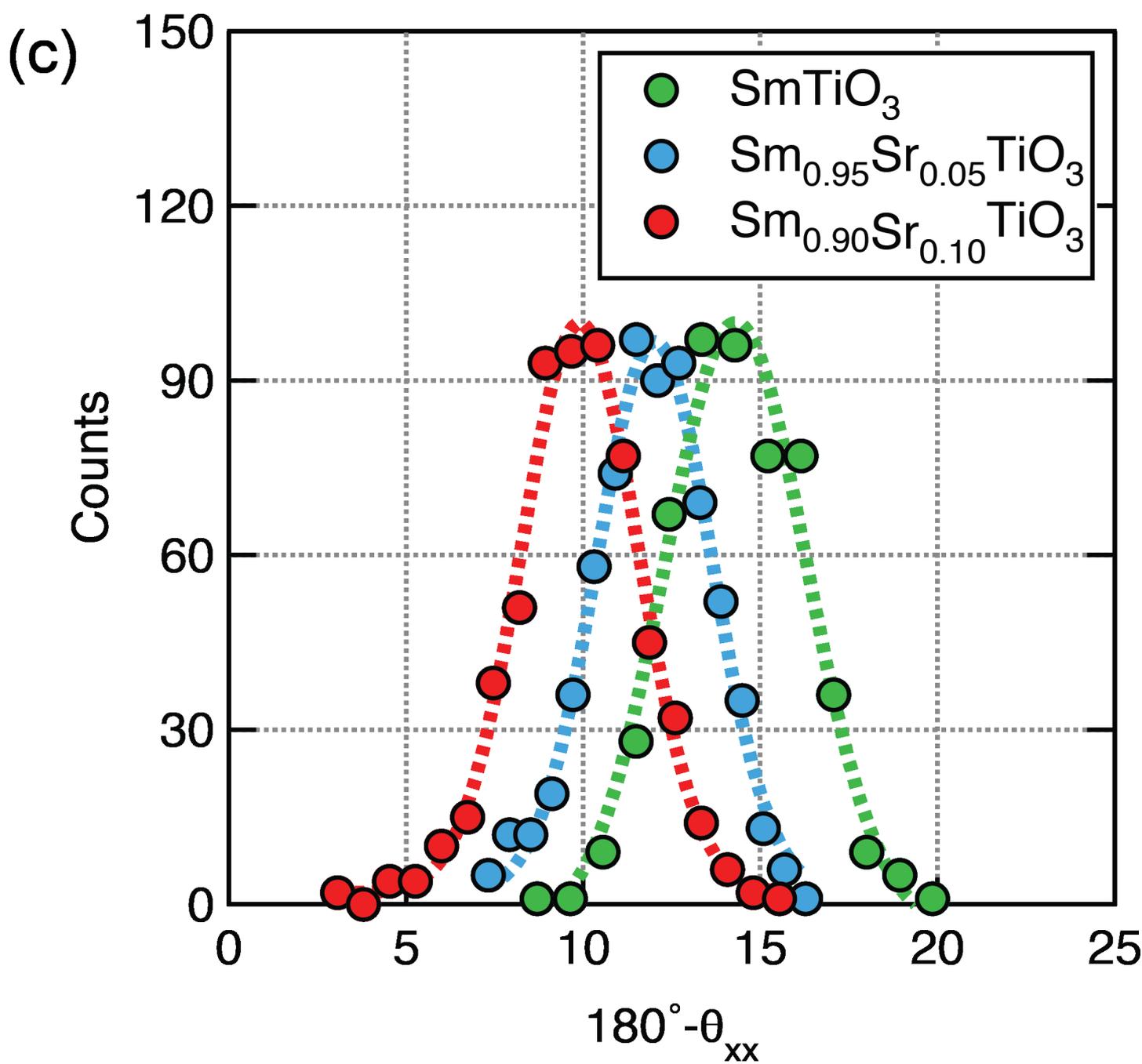

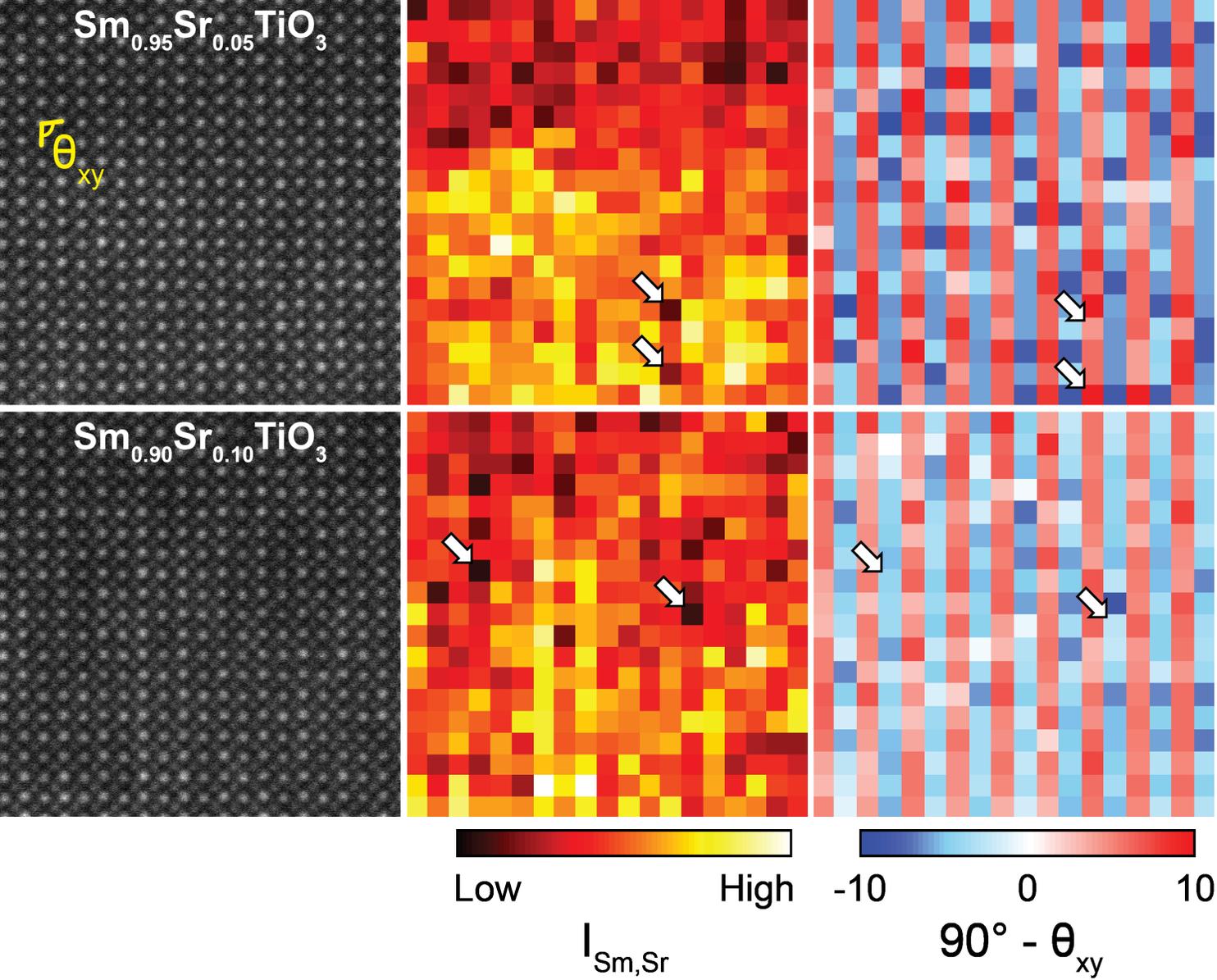